\documentclass[prd,twocolumn,floats,floatfix,nofootinbib]{revtex4}
\usepackage[dvips]{graphicx}
\usepackage{graphics}
\usepackage{dcolumn}
\usepackage{amssymb}
\usepackage{bm}
\usepackage{amsmath}
\usepackage{color}
\usepackage{epstopdf}

\def\spose#1{\hbox to 0pt{#1\hss}}

\def\lta{\mathrel{\spose{\lower 3pt\hbox{$\mathchar"218$}}
     \raise 2.0pt\hbox{$\mathchar"13C$}}}
\def\gta{\mathrel{\spose{\lower 3pt\hbox{$\mathchar"218$}}
     \raise 2.0pt\hbox{$\mathchar"13E$}}}
\newcommand{\be}{\begin{equation}}
\newcommand{\en}{\end{equation}}
\newcommand{\bea}{\begin{eqnarray}}
\newcommand{\ena}{\end{eqnarray}}

\begin{document}

\title{Hairy Black Holes from Horndeski Theory}

\author{ Santiago Esteban Perez Bergliaffa\footnote{sepbergliaffa@gmail.com}, Rodrigo Maier\footnote{rodrigo.maier@uerj.br},
Nathan de Oliveira Silvano\footnote{nathanosilvano@gmail.com} 
\vspace{0.5cm}}

\affiliation{Departamento de F\'isica Te\'orica, Instituto de F\'isica, Universidade do Estado do Rio de Janeiro,\\
Rua S\~ao Francisco Xavier 524, Maracan\~a,\\
CEP20550-900, Rio de Janeiro, Brazil\\
\\
}

\date{\today}

\begin{abstract}
We present an exact static
black hole
solution of Einstein field equations in the framework of
Horndeski Theory by imposing spherical symmetry and choosing the coupling constants
in the Lagrangian so that the only singularity in the solution is at $r=0$.
The analytical extension is built in two
particular domains of the parametric space. In the first domain we obtain a solution
exhibiting an event horizon analogous to that of the Schwarzschild geometry. For
the second domain, we show that the metric displays an exterior event horizon and
a Cauchy horizon which encloses a singularity. For both branches we obtain the corresponding Hawking temperature
which, when compared to that of the Schwarzschild black hole, acquires a 
correction proportional to a combination of the coupling constants.
Such a correction also 
modifies the definition of the entropy of the black hole.
\end{abstract}
\maketitle
\newpage
\section{Introduction}

The issue of late-time acceleration has motivated cosmologists to 
examine modifications of General Relativity
in the deep infrared (see \cite{clifton} and references therein). Although the cosmological constant is the simplest
candidate to drive the late-time acceleration, it poses an insurmountable problem to quantum field
theory, 
since it is hard to accommodate its observed value with vacuum energy calculations \cite{weinberg}.
Among several alternatives, Horndeski -- and beyond-Horndeski -- theories have been playing a significant role \cite{kase} in describing the accelerated expansion (see \cite{koba} for a review).
It is also of importance 
to study 
the effect of a Horndeski field 
on 
compact objects, such as neutron stars 
(see for instance \cite{masse})
and black holes. 
Over the last decade many examples of static and spherically symmetric hairy black holes in scalar-tensor theories
were obtained
in the literature. The simplest case in which solutions admit a hairy profile with a radially dependent
scalar field was 
studied 
in \cite{sotiriou1}-\cite{babichev}.
%
The case in which the scalar field depends linearly on time was considered in  \cite{edholm}-\cite{hamed} \footnote{For stability considerations about these models, see 
\cite{ogawa}-\cite{rham}.}. In fact, in \cite{khoury} the authors present a rather general analysis 
and obtain time-dependent
hairy black hole solutions within the Horndeski class of theories. Moreover, by considering 
perturbations of such solutions they show that fundamental instabilities are present, hence 
ruling out the
possibility of scalar hair with linear time
dependence in the Horndeski framework.
The authors of \cite{nicolis} provide a no-go theorem for static and sphericaly symmetric black hole solutions with vanishing-at-infinity galileon hair.  
They assume the following hypothesis: 
(i) asymptotic flatness, (ii) vanishing derivative of $\phi$ 
at infinity, (ii) the
finiteness of 
norm of the current $j^\mu$ (associated to shift invariance) 
down to the horizon, 
(iv) that the action includes a canonical kinetic term $\chi$, and (v) the functions $Q_i(\chi)$ (see Eq.\eqref{action})
are such that their $\chi$-derivatives contain only positive or zero
powers of $\chi$ as $\chi\rightarrow 0$ (\emph{i.e.} at infinity).
It is important to mention that this no-go theorem holds 
independently of the theory of gravitation
employed. 
Due to the fact that 
the radial component of the $4$-current $j^\mu$ -- the sole non-zero
component -- vanishes at the horizon (hence everywhere), 
it was shown in \cite{nicolis} that 
$\phi(r)={\rm constant}$ is the only possible solution.
Since the latter is equivalent to $\phi(r)=0$ due to the shift symmetry $\phi\rightarrow\phi+$const., 
spherically
symmetric black holes can only sustain trivial galileon profiles.

In \cite{babichev} the authors perform a 
detailed
analysis of the no-hair theorem examined in \cite{nicolis}
considering beyond-Horndeski theories, 
parameterized by 
six arbitrary functions of the kinetic term $\chi=-g^{\mu\nu} \phi_{,\mu} \phi_{\nu}/2$ --
four of them connected to ordinary Horndeski and two corresponding to the beyond-Horndeski case. Restricting 
to the case of ordinary Horndeski theory,
the general action presented in 
\cite{babichev}
can be written as
\begin{eqnarray}
\label{action}
\nonumber
S=\int \sqrt{-g}\Big\{Q_2(\chi)+Q_3(\chi)\Box\phi+ Q_4(\chi) R~~~~~~~~\\
\nonumber
+ Q_{4,\chi}[(\Box \phi)^2-(\nabla^\mu\nabla^\nu \phi)(\nabla_\mu\nabla_\nu \phi)]~~~~~~~~\\
+Q_5(\chi)G_{\mu\nu} \nabla^\mu\nabla^\nu \phi-\frac{1}{6}Q_{5,\chi}[(\Box \phi)^3~~~~~~~~~~~~~~\\
\nonumber
-3(\Box \phi) (\nabla^\mu\nabla^\nu \phi)(\nabla_\mu\nabla_\nu \phi)~~~~~~~~~~~~~~~~~~\\
\nonumber
+2(\nabla_\mu\nabla_\nu\phi)(\nabla^\nu\nabla^\gamma \phi)(\nabla_\gamma\nabla^\mu \phi)]
\Big\} d^4x.
\end{eqnarray}
Since it is important to check if the
hypotheses given above and considered by the no-hair theorem are indeed minimal,
here we aim at finding hairy black holes by 
lifting conditions (iii) and/or (iv).
In order to simplify the analysis, we shall restrict to the case
of a quartic Horndeski scalar field so that terms which contain $Q_5$ are absent.
In the next section we find an exact static and spherically symmetric solution of Einstein field equations assuming that 
$j^r$ (the radial component of the 4-current) vanishes at infinity and that the energy of the scalar field is finite
in 
a volume $V$ 
which encloses the region outside the external horizon
down to some value
where it is safe to assume that the Horndeski field behaves as in a Minkowski background.
In Section III the analytical extension of our solution is examined 
in two
particular domains of the parametric space. In the first domain we obtain a solution
exhibiting an event horizon analogous to that of the Schwarzschild geometry. For
the second domain, we show that the metric displays an outer event horizon and
a Cauchy horizon which encloses a singularity.
The corresponding Hawking temperatures are obtained in
section IV which is devoted to black thermodynamics.

\section{An Exact Solution}
Let us consider the action for the scalar field $\phi$ in Horndeski theory
\begin{eqnarray}
\label{eq0}
\nonumber
S=\int \sqrt{-g}\Big\{Q_2(\chi)+Q_3(\chi)\square\phi + Q_4(\chi)R\\
+Q_4,_\chi[(\square\phi)^2-(\nabla^\mu\nabla^\nu\phi)(\nabla_\mu\nabla_\nu\phi)]\Big\}d^4x,
\end{eqnarray}
where $g$ is the determinant of the metric and $R$ is the Ricci scalar built with the Christoffel symbols.

From the definition of the $4$-current, $$j^\nu=\frac{1}{\sqrt{-g}}\frac{\delta S}{\delta(\phi_{,\mu})},$$ it follows that 
\begin{eqnarray}
\label{eQ_2}
\nonumber
j^\nu=-Q_2,_\chi \phi^{, \nu}-Q_3,_\chi (\phi^{, \nu}\square\phi+\chi^{, \nu})~~~~~~~~~~~~~~~~~~~~\\
-Q_4,_\chi (\phi^{, \nu}R-2R^{\nu\sigma}\phi,_\sigma)~~~~~~~~~~~~~~~~~~~~~~~~~~~\\
\nonumber
-Q_4,_\chi,_\chi\{\phi^{, \nu}[(\square \phi)^2
-(\nabla_\alpha\nabla_\beta\phi)(\nabla^\alpha\nabla^\beta\phi)]\\
\nonumber
+2(\chi^{, \nu}\square\phi-\chi,_\mu\nabla^{\mu}\nabla^{\nu}\phi)
\},
\end{eqnarray} 
where we have used 
the usual convention for the Riemann tensor, 
\begin{eqnarray}
\label{eq1}
\nabla_\rho\nabla_\beta\nabla_\alpha\phi-\nabla_\beta\nabla_\rho\nabla_\alpha\phi=-R^\sigma_{~\alpha\rho\beta}\nabla_\sigma\phi.
\end{eqnarray}
Varying the action (\ref{eq0}) with respect to $g^{\mu\nu}$ we obtain the field equations
\begin{eqnarray}
\label{eq3}
Q_4 G_{\mu\nu}= T_{\mu\nu},
\end{eqnarray}
where
\begin{eqnarray}
\label{eq4}
\nonumber
T_{\mu\nu}=\frac{1}{2}(Q_2,_\chi \phi ,_\mu \phi ,_\nu+Q_2 g_{\mu\nu})
+\frac{1}{2}Q_3,_\chi(\phi ,_\mu \phi ,_\nu\square\phi~~~\\
\nonumber
-g_{\mu\nu} \chi,_\alpha \phi^{, \alpha}+\chi,_\mu \phi ,_\nu
+\chi,_\nu \phi ,_\mu)- Q_4,_\chi\Big\{\frac{1}{2}g_{\mu\nu}[(\square\phi)^2\\
\nonumber
-(\nabla_\alpha\nabla_\beta\phi)(\nabla^\alpha\nabla^\beta\phi)-2R_{\sigma\gamma}\phi^{,\sigma}\phi^{,\gamma}]
-\nabla_\mu\nabla_\nu \phi \square\phi\\
\nonumber
+\nabla_\gamma\nabla_\mu \phi \nabla^\gamma \nabla_\nu \phi-\frac{1}{2}\phi ,_\mu \phi ,_\nu R 
+R_{\sigma\mu}\phi^{,\sigma}\phi,_{\nu}\\
+R_{\sigma\nu}\phi^{,\sigma}\phi,_{\mu}+R_{\sigma\nu\gamma\mu} \phi^{,\sigma}\phi^{,\gamma}
\Big\}~~~~~~~~~~~~~~\\
\nonumber
-Q_4,_\chi,_\chi \Big\{g_{\mu\nu}(\chi,_{\alpha}\phi^{,\alpha}\square \phi+\chi_{,\alpha} \chi^{, \alpha})+\frac{1}{2}\phi ,_\mu \phi ,_\nu\times
\\
\nonumber
(\nabla_\alpha\nabla_\beta\phi\nabla^\alpha\nabla^\beta\phi-(\square\phi)^2)
- \chi,_\mu \chi,_\nu \\
\nonumber
- \square\phi( \chi,_\mu \phi,_\nu  
+ \chi,_\nu \phi,_\mu)
\\
\nonumber
- \chi,_\gamma[\phi^{,\gamma}\nabla_\mu\nabla_\nu\phi-(\nabla^\gamma\nabla_\mu\phi)\phi,_{\nu}
-(\nabla^\gamma\nabla_\nu\phi)\phi,_{\mu}]
  \Big\}.
\end{eqnarray} 
This expression
reduces to 
the correct one in the case of the canonical action for the scalar field. From now on we shall work with a scalar field $\phi\equiv\phi(r)$, which will be the source of a 
static and spherically symmetric geometry, described by 
the 
metric
%
\begin{eqnarray}
\label{eq16}
ds^2=-A(r)dt^2+\frac{1}{B(r)}dr^2+r^2(d\theta^2+\sin^2\theta d\varphi^2).
\end{eqnarray}
In this case, the only non-vanishing component of the $4$-current is given by
\begin{eqnarray}
\nonumber
j^r=-Q_2,_\chi B\phi^{\prime}-Q_3,_\chi\frac{(4A+r A^\prime)}{2rA}B^2{\phi^\prime}^2\\
+2Q_4,_\chi\frac{B}{r^2 A}[(B-1)A+r B A^\prime]\phi^\prime 
\\
\nonumber
- 2Q_4,_\chi,_\chi \frac{B^3(A+r A^\prime)}{r^2A}{\phi^\prime}^3,
\end{eqnarray}
where a prime denotes the derivative with respect to $r$. 
 
In order to simplify our analysis we now fix
\begin{eqnarray}
\label{eq5}
\nonumber
Q_2&=&\alpha_{21} \chi + \alpha_{22}{(-\chi)}^{\omega_2},\\
Q_3&=&\alpha_{31}{(-\chi)}^{\omega_3},\\
\nonumber
Q_4&=&\kappa^{-2} + \alpha_{42}{(-\chi)}^{\omega_4},
\end{eqnarray}
where $\kappa^2\equiv 8\pi G$.
Such choices generalize the ones proposed in 
\cite{babichev}. 
In the case of an
arbitrary static geometry 
given by Eq.(\ref{eq16}), we obtain 
\begin{eqnarray}
\label{eqjr}
\nonumber
j^r=\Gamma_1(r){\phi^\prime}+\Gamma_2(r){\phi^\prime}^{2\omega_2-1}~~~~~~~~~~~~~~~~~~~~~~~~\\
+\Gamma_3(r){\phi^\prime}^{2\omega_3}+\Gamma_4(r){\phi^\prime}^{2\omega_4-1},
\end{eqnarray}
where
\begin{eqnarray}
\nonumber
\Gamma_1(r)&=&-\alpha_{21}B,~~ \Gamma_2(r)=\frac{ \alpha_{22} \omega_2 B^{\omega_2}}{2^{\omega_2-1}},\\
\nonumber
\Gamma_3(r)&=&\frac{\alpha_{31} \omega_3 B^{\omega_3+1} (4 A + r A^\prime)}{2^{\omega_3}r A},\\
\nonumber
\Gamma_4(r)&=&\frac{\alpha_{42} \omega_4 B^{\omega_4}}{2^{\omega_4-2}r^2A}\{B(1-2\omega_4)(r A^\prime +A)+A\}.
\end{eqnarray}
It follows from Eq.(\ref{eqjr}) that
the trivial solution $\phi^{\prime} = 0$ might be avoided
if 
$\omega_2=1/2$, and/or $\omega_3=0$, and/or $\omega_4=1/2$. 
In fact, 
a quite simple hairy solution
was presented 
in \cite{babichev} 
for
the parameters
\begin{eqnarray}
\alpha_{22}=\alpha_{31}=0,~~\omega_4=\frac{1}{2}.
\end{eqnarray}
Such a solution 
displays a non-trivial
scalar field 
which generates a
geometry analogous to that described by the
Reissner-Nordstrom metric.

In order to set limits on the parameters $\omega_2$, $\omega_3$ and $\omega_4$,
let us examine the case of a Minkowski background
in spherical coordinates, so that
\begin{eqnarray}
\label{eq6}
ds^2=-dt^2+dr^2+
r^2(d\theta^2+\sin^2\theta d\varphi^2).
\end{eqnarray}
From Eq.(\ref{eQ_2}) the only non-vanishing component of the $4$-current is given by
\begin{eqnarray}
\label{eq7}
j^r=-Q_2,_\chi\phi^\prime-\frac{2Q_3,_\chi}{r}{\phi^\prime}^2-\frac{2Q_4,_\chi,_\chi}{r^2}{\phi^\prime}^3.
\end{eqnarray}
On the other hand, from Eq. (\ref{eq4}) we get
\begin{eqnarray}
\label{eq9}
\nonumber
T^0_{~~0}=\frac{Q_2}{2}+\frac{Q_3,_\chi {\phi^\prime}^2\phi^{\prime\prime}}{2}-\frac{Q_4,_\chi\phi^\prime}{r}\Big(\frac{\phi^\prime}
{r}+2\phi^{\prime\prime}      \Big)\\
+\frac{2Q_4,_\chi,_\chi{\phi^\prime}^3\phi^{\prime\prime}}{r}.
\end{eqnarray}
Using the \emph{Ansatz} given in Eq.(\ref{eq5}) we obtain
\begin{eqnarray}
\label{eq10}
\nonumber
j^r=-\alpha_{21}\phi^\prime+\frac{\alpha_{22}\omega_2{\phi^\prime}^{2\omega_2-1}}{2^{\omega_2-1}}+\frac{\alpha_{31}\omega_3{\phi^{\prime}}^{2\omega_3}}{2^{\omega_3-2}r}\\
+(1-\omega_4)\frac{\alpha_{42}\omega_4{\phi^\prime}^{2\omega_4-1}}{2^{\omega_4-3}r^2},
\end{eqnarray}
and
\begin{eqnarray}
\label{eq11}
\nonumber
T^0_{~~0}=-\Big(\frac{\alpha_{21}}{4}\Big){\phi^\prime}^2
+\frac{\alpha_{22}{\phi^\prime}^{2\omega_2}}{2^{\omega_2+1}}
+\frac{\alpha_{42}\omega_4{\phi^\prime}^{2\omega_4}}{2^{\omega_4-1}r^2}\\
-\Big[\frac{\alpha_{31}\omega_3{\phi^\prime}^{2\omega_3}}{2^{\omega_3}}
+(1-2\omega_4)\frac{\alpha_{42}\omega_4{\phi^\prime}^{2\omega_4-1}}{2^{\omega_4-2}r}\Big]\phi^{\prime\prime}.
\end{eqnarray}
For
$j^r$ to vanish at infinity, we assume that the derivative of the scalar field
behaves at infinity as %
\begin{eqnarray}
\label{eq12}
\lim_{r\rightarrow\infty}\phi^\prime\propto \frac{1}{r^p}. 
\end{eqnarray}
%
%
%
It follows from Eq.(\ref{eq10}) that 
in order to have
\begin{eqnarray}
\label{eq14}
\lim_{r\rightarrow\infty}j^r=0,
\end{eqnarray}
the following conditions are  sufficient
\footnote{We have also checked that these conditions are sufficient for the field given by 
$\phi ' \propto r^{-p}$ be a solution of the corresponding EOM at infinity.
}:
\begin{eqnarray}
\label{eq13a}
p>0,~~\omega_2 > \frac{1}{2},~~\omega_3 > -\frac{1}{2 p},~~\omega_4>\frac{1}{2}-\frac{1}{p}.
\end{eqnarray}

In order to set a more stringent limit on the parameters, let us consider the energy of the 
scalar field in a volume $V$ going from infinity down to some value of $r$, say $\bar{r}$, given by  
%
\begin{eqnarray}
\label{eq15}
E=\int_V\sqrt{-g}T^{0}_{~0}d^3x.
\end{eqnarray}
A necessary -- though not sufficient -- condition for the finiteness of the energy 
is that the integrand 
falls as $1/r^q$, with $q>1$, when $r \rightarrow \infty$. Hence 
we obtain the restriction $p>3/2$.
Taking into account this condition together with those given in Eq.(\ref{eq13a}), we proceed 
by setting
\begin{eqnarray}
\label{eqp2}
\alpha_{21}=\alpha_{31}=0,~~\omega_2=\frac{3}{2},~~\omega_4=\frac{1}{2},
\end{eqnarray}
in order to search for hairy configurations in Horndeski theory.
It is then easy to see that imposing $j^r=0$ one obtains
\begin{eqnarray}
\label{phi1}
\phi^{\prime}=\pm\frac{2}{r}\sqrt\frac{{-\alpha_{42}}}{{3B\alpha_{22}}}.
\end{eqnarray}
Notice that Eq. \eqref{phi1} leads to $\phi\propto \ln r$ for large $r$, but in principle this divergent behaviour is harmless because 
physical quantities
are defined in terms of the derivatives of $\phi$.

From the field equations it follows that
%
%
\begin{eqnarray}
\label{eqf}
A(R)=B(R)=1 -\frac{1}{R} + \frac{Q}{R}\ln{R},
\end{eqnarray}
where $R\equiv r/2GM$ and $Q\equiv q/2GM$, with
%
%
\begin{eqnarray}
\label{q}
q\equiv \Big(\frac{2}{3}\Big)^{3/2}\kappa^2 \alpha_{42}\sqrt{-\frac{\alpha_{42}}{\alpha_{22}}}.
\end{eqnarray}
%
%
Since
\begin{eqnarray}
\lim_{R\rightarrow\infty}A(R)=1,
\end{eqnarray}
the static solution given in Eq. (\ref{eqf}) is in fact asymptotically flat.
Finally, when $q\rightarrow 0$ we recover the Schwarzschild solution as 
expected. 

\begin{figure}[tbp]
\includegraphics[width=8cm,height=5cm]{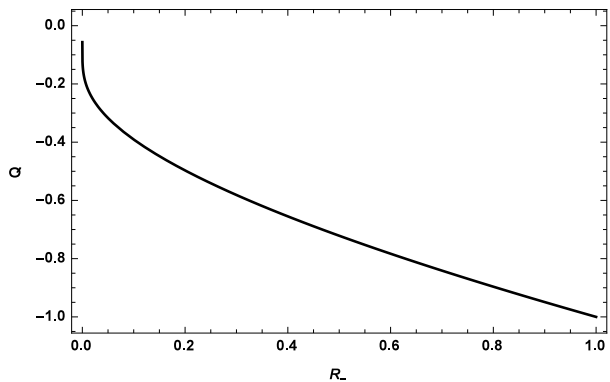}
\caption{Position of the internal horizon $R_-$ as a function of $Q$.}
\label{Rc}
\end{figure}

Let us next 
discuss 
whether, and under what circumstances, our configuration 
displays event horizons.
%
%
%
%
Firstly, it is easy to see that
\begin{eqnarray}
\label{ae1}
\lim_{R\rightarrow 0^+}B(R)=-{\rm sgn}(Q)\infty.
\end{eqnarray} 
On the other hand, one may show that $B(R)$ has only one local extremum at $R_e$ given by
\begin{eqnarray}
\label{ae2}
R_e=
\exp{\Big(1+\frac{1}{Q}}\Big).
\end{eqnarray}
These two results 
imply that there are two basic configurations of our solution:
\\
(i) If $Q>0$ we have only one 
event horizon $R_H$, such that $R_H=1$,  analogous to that of the Schwarzschild solution;
\\
(ii) if $Q<0$ 
the solution displays an inner horizon, located at $R=R_-(Q)$
(which, as as will be argued below,  
is a Cauchy horizon), and an outer horizon located at $R=1$. 
The position of the inner horizon as a function of $Q$ is shown in Fig.\ref{Rc}.
When $R_-\rightarrow R_+$ we obtain
the extremal configuration with an horizon $R_e=1$.
Hence the solution is somewhat analogous to the Reissner-Nordstr\"om spacetime.

Let us point out that 
we have checked that 
all the components of the Riemann tensor expressed in a convenient tetrad basis are finite outside and on the horizons. This is consistent with the finiteness of the components of the energy-momentum tensor calculated using Eq.\eqref{phi1}. Hence, the choice of parameters shown in Eq.\eqref{eqp2} leads to a 
solution that does not display the singularity present for instance in the Janis-Newman-Winicour spacetime \cite{janis}.

Before moving to the analytical extension, 
let us remark that 
our solution requires to set $\alpha_{21} = 0$ (see Eqns.\eqref{eq5} and \eqref{eqp2}). 
The ensuing equation of motion for the scalar field can be written in the form
(see for instance \cite{pappa2017})
\begin{eqnarray*}
\Box\phi+f(\partial\phi, \partial^2\phi)=0,
\end{eqnarray*}
where $f(\partial\phi, \partial^2\phi)$ is a nonlinear function of the first and second derivatives of $\phi$. 
The relevant problem of the propagation and well-posedness of initial data for the general Horndeski theory has been studied for instance in \cite{pappa2017a, pappa2017}, where it was shown that such a theory is weakly hyperbolic (in the weak-field limit)
assuming that $\alpha_{21}\neq 0$. The case of $\alpha_{21}=0$ will be studied in a future publication.

\section{Analytical Extension}

To proceed with the analytical extension of
the geometry described by Eq.(\ref{eqf}) it is useful to rescale the time 
coordinate
as $T=t/2GM$,
so that our solution can be rewritten in the form
\begin{eqnarray}
\label{ae4n}
\nonumber
ds^2=-B(R)dT^2+\frac{1}{B(R)}dR^2+R^2(d\theta^2+\sin^2{\theta}d\phi^2).
\end{eqnarray}
Let us then consider the following
coordinates transformation
\begin{eqnarray}
\label{ae3}
\frac{2\beta}{v}dv=\frac{1}{B(R)}dR+dT,\\
\label{ae31}
\frac{2\beta}{u}du=\frac{1}{B(R)}dR-dT,
\end{eqnarray}
where $\beta$ is an arbitrary constant. In this case, (\ref{ae4n}) reduces to 
\begin{eqnarray}
\label{ae4}
\nonumber
ds^2=\frac{4\beta^2 B(R)}{uv}dudv+R^2(d\theta^2+\sin^2{\theta}d\phi^2).
\end{eqnarray}
Defining
\begin{eqnarray}
\label{ae5}
R^\ast=\int\frac{1}{B(R)}dR,
\end{eqnarray}
a straightforward integration of Eqs. (\ref{ae3}) and (\ref{ae31}) furnishes
\begin{eqnarray}
\label{ae6}
R^\ast \equiv \beta \ln{|uv|},~~T\equiv \beta \ln{|v/u|}.
\end{eqnarray}

Let us now consider configuration (i) $(Q>0)$. In this case we have an event horizon $R_H$
analogous to that of the Schwarzschild solution. By considering the chart $(u_1, v_1)$
defined by $\beta>0$ we obtain,
\begin{eqnarray}
\label{ae7}
|u_1 v_1|=|R-R_H|^{h(R)/\beta} \sigma(R).
\end{eqnarray}
A straightforward integration of Eqs. (\ref{ae3}) and (\ref{ae31}) shows that $h(R)$ and $\sigma(R)$ are positive defined 
functions so that 
Eq.(\ref{ae7}) provides a regular covering for any subregion with 
$R > 0$.
A
Kruskal-type diagram is displayed in
Figure $4$. It gives a faithful map of any subregion covered by the chart
$(u_1, v_1)$. In this case, the maximal analytical extension of the spacetime is analogous
to that of a Schwarzschild black hole with an event horizon $R_H$ (cf. Fig. $5$).
\begin{figure}[tbp]
\includegraphics[width=7cm,height=4cm]{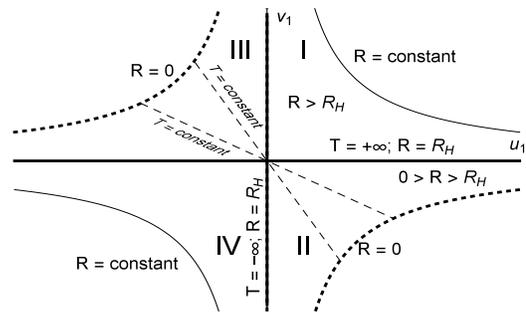}
\caption{The Kruskal extension of the spacetime for $Q>0$.}
\end{figure}
\begin{figure}[tbp]
\includegraphics[width=7cm,height=4cm]{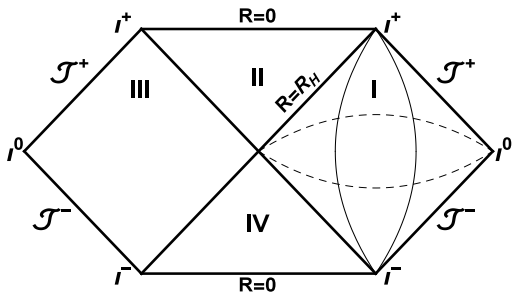}
\caption{Penrose diagram for the analytical extension with $Q>0$. Thin solid and dashed lines are connected
to $R={\rm constant}$ and $T={\rm constant}$, respectively.}
\end{figure}

Taking into account the cosmic censorship conjecture \cite{penrose}, we shall restrict to the case
in which there is at least one 
event horizon 
in configuration (ii) $(Q<0)$.
In this case it can be easily seen that
if $\beta>0$ one may consider the chart $(\bar{u}_1, \bar{v}_1)$ for $R > R_-$ defined as
\begin{eqnarray}
\label{ae8}
|\bar{u}_1 \bar{v}_1|=\Big|\frac{R-R_+}{R-R_-}\Big|^{\bar{h}(R)/[\beta(R_+-R_-)]} \bar{\sigma}(R).
\end{eqnarray}
Again, a straightforward integration of Eqs. (\ref{ae3}) and (\ref{ae31}) shows that $h(R)$ and $\sigma(R)$ are positive defined 
functions.
Therefore,
the metric develops no coordinate singularity at $R_+$. However, this chart
does not furnish a regular covering for any subregion with $R < R_+$. In order to circumvent
this problem, one may resort again to the chart given by Eq.(\ref{ae8}) but this time with $\beta<0$, so that the metric develops no coordinate singularity at $R_-$.
In fact, 
such a chart gives a regular
map of any given subregion of the manifold with 
$R < R_+$. In the domain 
$R_-<R<R_+$ the two charts overlap furnishing a regular
map for the entire manifold.
In Fig. 6 we show the
main part of the Penrose diagram obtained from this maximal analytical extension. 
It is analogous to that of Reissner-Nordstrom
spacetime, exhibiting an exterior event
horizon $R_+$ together with an interior Cauchy horizon \footnote{It can be seen from the figure that the evolution of data given on an initial timelike hypersurface outside the black hole is non-unique inside $R_-$, since it can be affected by data propagating from the singularity at $r=0$ \cite{burko}.}
$R_-$,
which encloses a
physical singularity at $R = 0$. For $R_+= R_-=R_e$ analogous calculations provide the maximal analytical extension for the extremal configuration.
\begin{figure}[tbp]
\includegraphics[width=10cm,height=6cm]{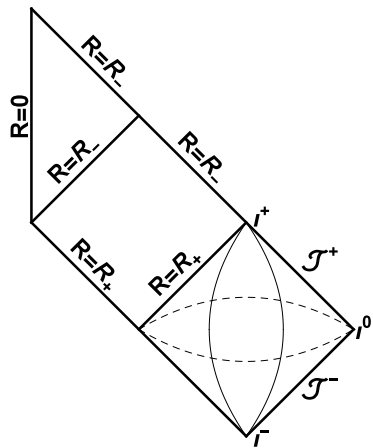}
\caption{Penrose diagram for configuration for $Q<0$. According to the cosmic censorship conjecture, we are restricting to the case
in which an event horizon $R_+$ is formed. 
Here we see that the
spacetime exhibits an exterior event horizon $R_+$ and an interior Cauchy horizon $R_-$. 
The complete analytical extension can be obtained by connecting asymptotically
flat regions, like the fundamental portion shown above, in an infinite chain.}
\end{figure}

\section{Black Hole Thermodynamics}

Using a semi-classical approach, S. W. Hawking derived the thermal spectrum of emitted
particles by a black hole \cite{hawking}. Following his same original procedure, we now intend to
obtain the temperature of Hawking radiation in the context of Eq.(\ref{eqf}). To do so, let us then consider a test 
massless Klein-Gordon field $\psi$ in such a background. 
The propagation of the scalar field
is then described by the scalar wave equation
\begin{eqnarray}
\label{bt1}
\Box \psi=0.
\end{eqnarray}
Given the symmetries of the background we search for a solution as
\begin{eqnarray}
\label{bt2}
\psi_{\omega m l}=\frac{1}{R}\Delta_{\omega l}(R^\ast)Y_{ml}(\theta, \varphi)e^{-i\omega t},
\end{eqnarray}
so that the wave equation (\ref{bt1}) reduces to an ordinary differential
equation in $R^\ast$. In fact, it is straightforward to see that in the asymptotical limit $R \rightarrow \infty$,
this equation reduces to
\begin{eqnarray}
\label{bt3}
\frac{d^2\Delta_{\omega l}}{d{R^\ast}^2}+\omega^2 \Delta_{\omega l}=0 \rightarrow \Delta_{\omega l}(R^\ast)=e^{\pm i\omega R^\ast}.
\end{eqnarray}
Therefore one may write the two branches of the asymptotic Klein Gordon field as
\begin{eqnarray}
\label{bt4}
\psi_U=\frac{1}{R}e^{-i\omega U}Y_{ml}(\theta, \varphi),\\
\label{bt41}
\psi_V=\frac{1}{R}e^{-i\omega V}Y_{ml}(\theta, \varphi),
\end{eqnarray}
where $U$ and $V$ are null coordinates defined as
\begin{eqnarray}
V:=T+R^\ast,~~ U:=T-R^\ast. 
\end{eqnarray}

Let us now assume that the source that generates the exterior solution is given
by a thin shell of a spherically symmetric matter distribution, and the flat spacetime
inside such distribution is given by
\begin{eqnarray}
\label{bt5}
ds^2=-d\bar{t}^2+d\bar{r}^2+\bar{r}^2(d\bar{\theta}^2+\sin^2{\bar{\theta}}d\bar{\varphi}^2).
\end{eqnarray}
Defining $a(\bar{t})$ as the scale factor that describes the evolution of the matter
distribution, we impose that the interior metric matches the exterior geometry by the
following equation
\begin{eqnarray}
\label{bt6}
-1+\Big(\frac{da}{d\bar{t}}\Big)^2=-B(a)\Big(\frac{dT}{d\bar{t}}\Big)^2+\frac{1}{B(a)}\Big(\frac{da}{d\bar{t}}\Big)^2.
\end{eqnarray}
It is well know that the temperature of a Schwarzschild black hole is quite different from the Reissner-Nordstrom
solution, although both depend of the radius of the event horizon. In fact, in the case of a Reissner-Nordstrom black hole, it 
can be seen that its temperature 
vanishes as one approaches to the extremal configuration. As in this section we intend to evaluate both temperatures
corresponding to configurations with $Q>0$ and $Q<0$, from now on we denote the event horizons of both configurations
by the same variable $R_H$ (although they are different in each case). 

Assuming that the null incident rays reach the matter distribution at a given time $\bar{t}_i$
with $a(\bar{t}_i) \equiv a_i \gg R_H$, 
we obtain that $A(r) \sim 1$ and from (\ref{bt6})
\begin{eqnarray}
\label{bt7}
\Big(\frac{dT}{d\bar{t}}\Big)^2 \simeq 1 \rightarrow T \simeq \bar{t}.
\end{eqnarray}
By defining the respective null interior coordinates by
\begin{eqnarray}
\bar{V}=\bar{t}+\bar{r},~~\bar{U}=\bar{t}-\bar{r}.
\end{eqnarray}
we end up with the following relations
\begin{eqnarray}
\label{bt8}
V_i\simeq \bar{t}+R^\ast \rightarrow \bar{V}_i=V_i+\Omega_i.
\end{eqnarray}
where $\Omega_i\equiv a_i-R^\ast(a_i)$.

When $\bar{r}=0$, we derive the trivial relation between $\bar{V}$ and $\bar{U}$ at the center of the
matter distribution:
\begin{eqnarray}
\label{bt9}
\bar{V}_0=\bar{U}_0.
\end{eqnarray}

Let us now consider that the outgoing waves emerge from the matter distribution when
$a(\bar{t})\sim R_H$. If $\bar{t}_{e}$ is taken to be the instant in 
which $a(\bar{t}_{e})=R_H$, one may expand
the scale factor $a({\bar{t}})$ in Taylor series as
\begin{eqnarray}
\label{bt10}
a(\bar{t})=R_H+\dot{a}_e(\bar{t}_e-\bar{t}).
\end{eqnarray}
where $\dot{a}_e \equiv \frac{da}{d\bar{t}}\Big|_{\bar{t}_e}$.
Therefore, up to first order in $(\bar{t}_e-\bar{t})$ from equation (\ref{bt6}) we have
\begin{eqnarray}
\label{bt11}
T\simeq - \zeta \ln{|\bar{t}_e-\bar{t}|}
\end{eqnarray}
where
\begin{eqnarray}
\label{bt12}
\zeta := \Big(\frac{dB}{dR}\Big)^{-1}\Big|_{R_H} \equiv \frac{R_H^2}{R_H+Q}.
\end{eqnarray}

At this stage, it is worth mentioning that both configurations (with $Q>0$ or $Q<0$) yield similar results for $r^\ast(r\sim r_H)$,
namely, 
\begin{eqnarray}
\label{bt13}
R^\ast \simeq \zeta\ln\Big|\frac{R}{R_H}-1\Big|
\end{eqnarray}
Therefore we obtain
\begin{eqnarray}
\label{bt14}
U_{e}\simeq - 2\zeta \ln{|\bar{t}_e-\bar{t}|}
\end{eqnarray}
for both configurations so that
\begin{eqnarray}
\label{bt15}
\bar{U}_{e}\propto \exp{\Big[-\frac{U_{e}}{2\zeta}\Big]}
\end{eqnarray}

At the origin of the coordinate system we have $\bar{U}_0 = \bar{V}_0$. Therefore the
relation between the exterior null coordinates is given by
\begin{eqnarray}
\label{bt17}
V\propto \exp{\Big[-\frac{U}{2\zeta}\Big]}\rightarrow U=-2\zeta\ln{V}
\end{eqnarray}

From (\ref{bt4}) one may now expand $\psi_{U}$ in terms of $\psi_{V}$ as
\begin{eqnarray}
\label{bt18}
\nonumber
\psi_U=\int_0^\infty[\gamma^\ast_{\omega^\prime \omega m l}\exp{(-i\omega^\prime V)}+\beta_{\omega^\prime \omega m l}\exp{(i\omega^\prime V)}]d\omega^\prime.
\end{eqnarray}
where $\gamma^\ast_{\omega^\prime \omega m l}$ and $\beta_{\omega^\prime \omega m l}$ are the Bogolubov coefficients\cite{bogolubov}. Therefore, it is straightforward to show\cite{hawking} that
\begin{eqnarray}
\label{bt19}
|\gamma_{\omega^\prime \omega m l}|=e^{2\pi\omega \zeta} |\beta_{\omega^\prime \omega m l}|.
\end{eqnarray}

On the other hand, it follows from the orthogonality propriety of $\psi_U$ and $\psi_V$ that
\begin{eqnarray}
\label{bt20}
\sum_{\omega^\prime}[|\gamma_{\omega^\prime \omega m l}|^2-|\beta_{\omega^\prime \omega m l}|^2]=1.
\end{eqnarray}
Therefore we obtain that the spectrum of the average number of created particles
on the $\omega m l$ mode is given by
\begin{eqnarray}
\label{bt21}
N_{\omega m l}=\sum_{\omega^\prime}|\beta_{\omega^\prime \omega m l}|^2=\frac{1}{e^{4\pi\omega\zeta}-1}.
\end{eqnarray}
At this stage we draw the reader's attention to a word to note. As we working in rescaled units $T=t/2GM$,
the frequency appearing in (\ref{bt21}) is also rescaled by a factor 2GM. Restoring this factor
the above result corresponds to a Planckian spectrum with associated temperature
\begin{eqnarray}
\label{bt22}
T_{H}=\frac{1}{8\pi GM\zeta},
\end{eqnarray}
which,
using Eq.(\ref{bt12}) and the fact that both for positive and negative $Q$, the relevant value of $R_H$ is 1,
amounts to
\begin{eqnarray}
\label{bt23}
T_{H}=\frac{1+Q}{8\pi GM}.
\end{eqnarray}
In the limit $Q\rightarrow 0$ we recover the temperature corresponding to the Schwarzschild black hole, namely $T_H=1/8\pi G M$. In our solution, the temperature acquires a correction which depends of the parameters of the theory, causing it to be higher or lower than that of the Scharzschild case, depending on the sign of $Q$.

In the case of $Q<0$, 
the temperature can be conveniently rewritten using the relation 
%
\begin{eqnarray}
\label{bt24}
Q \equiv \frac{1-R_-}{\ln(R_-)},
\end{eqnarray}
obtained
from the condition $B(R_H)=0=B(R_-)$.
It follows that
%
%
%
\begin{eqnarray}
\label{bt26}
\nonumber
T_H=\frac{1}{8\pi GM }+\Big(\frac{1-R_-}{8\pi GM }\Big) \times 
\frac{1}{\ln(R_-)}
.
\end{eqnarray}
Hence, in the limit of extremal configurations, namely $R_- \rightarrow R_H$, we obtain that 
$T_H\rightarrow 0$ as expected.

To conclude this section we now refer to the analysis of energy processes involving black holes.
It is well known \cite{beken} that the entropy of a black hole should be proportional
to the area of its event horizon. According to Bekenstein First Law of Black
Hole Thermodynamics, the surface gravity of the black hole appears as proportional
(via dimensional fundamental constants) to a temperature. Bekenstein's
results however are rather geometrical and do not involve any fundamental principle of statistical mechanics.
In fact, S. W. Hawking\cite{hawking} showed that the black holes should emit particles
with a Planckian thermal spectrum of temperature $T_H = \kappa /2\pi$ (in natural units
$\hslash = c = G = K_B = 1$) where $\kappa = 1/4M$ is its surface gravity. 
This striking result fits exactly in the Bekenstein formula thus validating his proposals and fixing the
proportionality factor connecting the entropy and the area of the black hole.

In order to extend this original result, it is useful consider our solution in the form (\ref{eqf})
so that
\begin{eqnarray}
\label{bt27}
B(r_H)=1 -\frac{2GM}{r_H} + \frac{q}{r_H}\ln{\Big(\frac{r_H}{{2GM}}\Big)}\equiv 0.
\end{eqnarray}
The exact differential of the above equation is given by
\begin{eqnarray}
\label{bt28}
r_H dr_H={ 4G^2 M\zeta (1+Q)dM}.
\end{eqnarray}
By defining the the outer horizon area as $A_{out} := 4\pi r_H^2$
we then obtain that
\begin{eqnarray}
\label{bt29}
\nonumber
d A_{out} &=&  8\pi r_H dr_H= \frac{4G}{T_H}{(1+Q)} dM.
\end{eqnarray}
We can therefore associate the horizon area of the black hole with the geometrical entropy
\begin{eqnarray}
\label{bt30}
S_{geom}=\frac{1}{4{(1+Q)}}A_{out},
\end{eqnarray}
which shows 
the form of
Bekenstein's first law of black hole thermodynamics \cite{beken} is valid for our solution.

\section{Final Remarks}

We presented a new static solution of Einstein field equations
in the framework of Horndeski theory. Choosing a suitable domain for the coupling constants, we 
obtained  asymptotically flat configurations which satisfy the cosmic censorship hypothesis \cite{penrose}
so that naked singularities are not  considered. 
 It is shown that the
structure of spacetime bifurcates in two main particular branches. In the first domain,
the analytical extension is analogous to that of the Schwarzschild solution with an event
horizon. For the second branch, we show that the geometry provides an exterior event
horizon and an interior (Cauchy) horizon which encloses the singularity. 

The goal of the analysis performed in this paper was to seek for static 
solutions in Horndeski theory in order to better understand how such degrees
of freedom may provide deviations from General Relativity. In this sense, our main
result shows how the presence of a galileon field leads to 
a temperature that is different from that of the Schwarzschild black hole, the difference being dependent of the coupling constants of the theory. 
Furthermore, we also examine the effects of the scalar field on Hawking radiation\cite{hawking} as a simple
application. In this case the calculation of the modified
Hawking temperature allowed us to derive, analogously to Bekenstein\cite{beken}, a geometric
entropy that confirms the classical prediction that the entropy is proportional to the
area of the event horizon, and is corrected by $Q$. Although the classical black hole thermodynamics introduced
by Bekenstein was validated by Hawking's semiclassical derivation of the black body
thermal emission of a black hole, black hole thermodynamics always seemed to possess
a heuristic character since no basic principle of statistical mechanics was used in its
derivation. Indeed, the definition of the entropy of black holes is still an open issue
and we actually refer to it as a geometrical entropy.

Some features of the solution presented in Eq.(\ref{eqf})
still need to be scrutinized.
For instance, the presence of $R_{-}$ 
poses the question of the stability of such a spacetime. Due to its similarity to the Reissner-Nordstr\"om
spacetime, we should expect that the flux of energy of test fields may diverge on crossing
$R_{-}$. However, a complete treatment of the problem should include higher order nonlinear terms 
so as to provide sufficient conditions
for instability.
It would also be of interest to determine the internal solution generated by
a fluid distribution which could be matched, hence engender, the external solution. 
Last but not least, an extension of the field represented by Eq. (\ref{eqf}) by including a non-trivial time dependence
is worthy of analysis. In this case, perturbations 
could establish if 
such solution
suffers from fatal instabilities as those shown in \cite{khoury}.
We hope to address these points in a future publication.

\end{document}